\documentclass[12pt]{article}
\usepackage{amsmath,amssymb,amsthm,amsxtra,overpic,bbm,bm,epsfig,ulem}
\usepackage{color}
\usepackage{cite}

\usepackage{hyperref}
\usepackage{url}

\def\thefootnote{\fnsymbol{footnote}}

\begin{document}

\vspace{0.2cm}

\begin{center}
{\large\bf New rephasing invariants and CP violation built from the trios \\
of the CKM or PMNS matrix elements}
\end{center}

\vspace{0.2cm}

\begin{center}
{\bf Shu Luo}$^1$
\footnote{E-mail: luoshu@xmu.edu.cn}
and
{\bf Zhi-zhong Xing$^{2,3,4}$}
\footnote{E-mail: xingzz@ihep.ac.cn}
\\
{\small $^{1}$Department of Astronomy, Xiamen University, Fujian 361005, China} \\
{\small $^{2}$Institute of High Energy Physics, Chinese Academy of Sciences,
Beijing 100049, China} \\
{\small $^{3}$School of Physical Sciences,
University of Chinese Academy of Sciences, Beijing 100049, China} \\
{$^{4}$Center of High Energy Physics, Peking University, Beijing 100871, China}
\end{center}

\vspace{1cm}

\begin{abstract}
Given the $3\times 3$ Cabibbo-Kobayashi-Maskawa (CKM) quark flavor mixing
matrix $V$, we define a new set of rephasing invariants in terms of the
``trios" of its nine elements: $\lozenge^{ijk}_{\alpha\beta\gamma} \equiv
(V^{}_{\alpha i} V^{}_{\beta j} V^{}_{\gamma k})/\det V$ with
$\alpha \neq \beta \neq \gamma$ and $i \neq j \neq k$ running respectively
over $(u, c, t)$ and $(d, s, b)$. We find that ${\rm Im}
\lozenge^{ijk}_{\alpha\beta\gamma} = - {\cal J}$ holds, where ${\cal J}$
is the well-known Jarlskog invariant of weak CP violation. Analogous rephasing
invariants $\blacklozenge^{ijk}_{\alpha\beta\gamma} \equiv (U^{}_{\alpha i}
U^{}_{\beta j} U^{}_{\gamma k})/\det U$ can be defined for the $3\times 3$
Pontecorvo-Maki-Nakagawa-Sakata (PMNS) lepton flavor mixing matrix $U$,
where $\alpha \neq \beta \neq \gamma$ and $i \neq j \neq k$ run respectively
over $(e, \mu, \tau)$ and $(1, 2, 3)$. Taking into account small non-unitarity
of $U$ based on the canonical seesaw mechanism for neutrino mass generation,
we calculate ${\rm Im} \blacklozenge^{ijk}_{\alpha\beta\gamma}$ with the help
of a full Euler-like block parametrization of the seesaw flavor structure and
demonstrate that their leading terms converge to a universal
invariant ${\cal J}^{}_\nu$ in the unitarity limit of $U$.
\end{abstract}

\newpage

\def\thefootnote{\arabic{footnote}}
\setcounter{footnote}{0}

\section{Motivation}

Within the standard model (SM) of electroweak interactions, it is the
$3\times 3$
Cabibbo-Kobayashi-Maskawa (CKM) matrix that describes flavor mixing and
CP violation in the quark sector~\cite{Cabibbo:1963yz,Kobayashi:1973fv}:
\begin{eqnarray}
-{\cal L}^{}_{\rm cc} = \frac{g}{\sqrt{2}} \hspace{0.1cm}
\overline{\big(\begin{matrix} u & c & t \end{matrix}\big)^{}_{\rm L}}
\hspace{0.1cm} \gamma^\mu
\left(\begin{matrix}
V^{}_{ud} & V^{}_{us} & V^{}_{ub} \cr
V^{}_{cd} & V^{}_{cs} & V^{}_{cb} \cr
V^{}_{td} & V^{}_{ts} & V^{}_{tb} \cr\end{matrix}\right) \hspace{-0.1cm}
\left(\begin{matrix}
d \cr s \cr b \cr\end{matrix}\right)_{\rm L} \hspace{-0.1cm} W^+_\mu
+ {\rm h.c.} \; ,
\label{1}
\end{eqnarray}
where the elements of $V$ involve an {\it irremovable} phase which
violates the CP symmetry of weak interactions. One may identify this
nontrivial Kobayashi-Maskawa (KM) phase~\cite{Kobayashi:1973fv} either
in a specific parametrization of $V$, such as the one advocated by
the Particle Data Group~\cite{ParticleDataGroup:2024cfk},
\begin{eqnarray}
V = \left( \begin{matrix} c^{}_{12} c^{}_{13} & s^{}_{12}
c^{}_{13} & \hat{s}^*_{13} \cr
-s^{}_{12} c^{}_{23} - c^{}_{12} \hat{s}^{}_{13} s^{}_{23} &
c^{}_{12} c^{}_{23} - s^{}_{12} \hat{s}^{}_{13} s^{}_{23} &
c^{}_{13} s^{}_{23} \cr
s^{}_{12} s^{}_{23} - c^{}_{12} \hat{s}^{}_{13} c^{}_{23} &
-c^{}_{12} s^{}_{23} - s^{}_{12} \hat{s}^{}_{13} c^{}_{23} &
c^{}_{13} c^{}_{23}
\cr \end{matrix} \right) \;
\label{2}
\end{eqnarray}
with $c^{}_{ij} \equiv \cos\vartheta^{}_{ij}$,
$s^{}_{ij} \equiv \sin\vartheta^{}_{ij}$
and $\hat{s}^{}_{13} \equiv s^{}_{13} e^{{\rm i}\delta^{}_{\rm KM}}$
(for $ij = 12, 13, 23$); or in a rephasing-invariant approach with
the help of a ``quartet" of the CKM matrix elements involving four
different quark flavor indices, such as~\cite{ParticleDataGroup:2024cfk}
\footnote{Historically, $\beta^{}_{\rm KM}$ was the first physical phase
parameter that was experimentally determined in a clean
way~\cite{BaBar:2001pki,Belle:2001zzw} and hence served as a smoking gun
of the KM mechanism of CP violation.}
\begin{eqnarray}
\beta^{}_{\rm KM} \equiv \arg\left(-\frac{V^{}_{cd} V^*_{cb}}{V^{}_{td}
V^*_{tb}}\right) \; ,
\label{3}
\end{eqnarray}
which is insensitive to redefining the phases of quark fields.
Of course, $\sin\delta^{}_{\rm KM}$ and $\sin\beta^{}_{\rm KM}$ must
be proportional to each other, as their physical meanings are
substantially the same.

But the Jarlskog invariant $\cal J$ is known as the unique quantity
that characterizes the strength of weak CP violation in the
SM~\cite{Jarlskog:1985ht,Wu:1985ea},
\begin{eqnarray}
{\rm Im}\left(V^{}_{\alpha i} V^{}_{\beta j} V^{*}_{\alpha j}
V^{*}_{\beta i} \right) =
{\cal J} \sum_\gamma \epsilon^{}_{\alpha\beta\gamma} \sum_{k}
\epsilon^{}_{ijk} \; ,
\label{4}
\end{eqnarray}
where $\epsilon^{}_{\alpha\beta\gamma}$ and $\epsilon^{}_{ijk}$ denote the
three-dimensional Levi-Civita symbols, and the Greek and Latin subscripts run
respectively over $(u, c, t)$ and $(d, s, b)$. It is actually the unitarity
of the $3\times 3$ CKM matrix that assures $\cal J$ to be a universal
rephasing-invariant measure of CP violation~\cite{Xing:2020ijf}.
Given the parametrization of $V$ in Eq.~(\ref{2}), for example,
we are left with ${\cal J} = c^{}_{12} s^{}_{12} c^2_{13} s^{}_{13}
c^{}_{23} s^{}_{23} \sin\delta^{}_{\rm KM}$. A global analysis of the
currently available experimental data on quark flavor mixing and CP
violation allows one to determine $\cal J$ to a
good degree of accuracy: ${\cal J} = \left(3.12^{+0.13}_{-0.12}\right)\times
10^{-5}$~\cite{ParticleDataGroup:2024cfk}
\footnote{In particular, a successful numerical determination of $\cal J$
by using four independent moduli of the CKM matrix elements turns out to
be feasible today~\cite{Luo:2023fcc}.}.

The first purpose of this work is to propose a new set of rephasing
invariants built from the ``trios" of the CKM matrix elements instead
of their ``quartets":
\begin{eqnarray}
\lozenge^{ijk}_{\alpha\beta\gamma} \equiv \frac{1}{\det V} \hspace{0.05cm}
\big(V^{}_{\alpha i} V^{}_{\beta j} V^{}_{\gamma k}\big) \; ,
\label{5}
\end{eqnarray}
where the Greek and Latin indices $\alpha \neq \beta \neq \gamma$ and
$i \neq j \neq k$ run respectively over $(u, c, t)$
and $(d, s, b)$, and $\det V$ is the determinant of $V$.
We are going to show that ${\rm Im} \lozenge^{ijk}_{\alpha\beta\gamma}
= - {\cal J}$ holds.

It is certainly straightforward to extend the new rephasing invariants
$\lozenge^{ijk}_{\alpha\beta\gamma}$ in the quark sector to those in
the lepton sector beyond the SM, after nonzero neutrino masses and lepton
flavor mixing effects are taken into account. Here let us concentrate on
the canonical seesaw mechanism~\cite{Minkowski:1977sc,Yanagida:1979as,
GellMann:1980vs,Glashow:1979nm,Mohapatra:1979ia}, which has widely
been accepted as a most natural and economical extension of the SM
to understand the origin of tiny neutrino masses. Then lepton flavor
mixing and CP violation in weak charged-current interactions are
described by the $3 \times 3$ Pontecorvo-Maki-Nakagawa-Sakata (PMNS)
matrix $U$~\cite{Pontecorvo:1957cp,Maki:1962mu,Pontecorvo:1967fh}
associated with three light Majorana neutrinos and its counterpart
$R$ associated with three heavy Majorana neutrinos:
\begin{eqnarray}
-{\cal L}^{\prime}_{\rm cc} = \frac{g}{\sqrt{2}} \hspace{0.1cm}
\overline{\big(\begin{matrix} e & \mu & \tau\end{matrix}\big)^{}_{\rm L}}
\hspace{0.1cm} \gamma^\mu \left[ U \left( \begin{matrix} \nu^{}_{1}
\cr \nu^{}_{2} \cr \nu^{}_{3} \cr\end{matrix}
\right)^{}_{\hspace{-0.08cm} \rm L}
+ R \left(\begin{matrix} N^{}_4 \cr N^{}_5 \cr N^{}_6
\cr\end{matrix}\right)^{}_{\hspace{-0.08cm} \rm L} \hspace{0.05cm} \right]
W^-_\mu + {\rm h.c.} \; ,
\label{6}
\end{eqnarray}
where $U$ and $R$ are correlated with each other via both the unitarity
condition $U U^\dagger + R R^\dagger = {\bf 1}$ and the exact
seesaw relation $U D^{}_\nu U^T + R D^{}_N R^T = {\bf 0}$ with
$D^{}_\nu = {\rm Diag}\{m^{}_1, m^{}_2, m^{}_3\}$ and $D^{}_N =
{\rm Diag}\{M^{}_4, M^{}_5, M^{}_6\}$ being the diagonal mass
matrices~\cite{Xing:2007zj,Xing:2011ur}. The second purpose of
this paper is to introduce the new leptonic rephasing invariants
\begin{eqnarray}
\blacklozenge^{ijk}_{\alpha\beta\gamma} \equiv \frac{1}{\det U}
\hspace{0.05cm}
\big(U^{}_{\alpha i} U^{}_{\beta j} U^{}_{\gamma k}\big) \; ,
\label{7}
\end{eqnarray}
where the Greek and Latin indices $\alpha \neq \beta \neq \gamma$ and
$i \neq j \neq k$ run respectively over
$(e, \mu, \tau)$ and $(1, 2, 3)$, and $\det U$ denotes the
determinant of $U$. Given the fact that all the nine active-sterile
flavor mixing angles of $R$ are tiny thanks to the highly suppressed
non-unitarity effects of $U$~\cite{Blennow:2023mqx,Xing:2024gmy},
we are going to demonstrate that the leading terms of
$\blacklozenge^{ijk}_{\alpha\beta\gamma}$ converge to a universal
quantity --- the leptonic Jarlskog invariant ${\cal J}^{}_\nu$
in the unitarity limit of the PMNS matrix $U$.

\section{On the CKM trios}

Compared with the conventional rephasing invariants built from the
``quartets" of the nine CKM matrix elements~\cite{Jarlskog:1985ht,Wu:1985ea},
the new ``trio"-style invariants defined in Eq.~(\ref{5}) involve all
the six quark flavor indices and thus must be CP-violating. To see why
$\lozenge^{ijk}_{\alpha\beta\gamma}$ are independent of any redefinitions
of the phases of six quark fields, let us consider the arbitrary phase
transformations
\begin{eqnarray}
q^{}_\alpha \to q^{}_\alpha e^{{\rm i} \phi^{}_\alpha} \; , &&
q^{}_i \to q^{}_i e^{{\rm i}\phi^{}_i} \; ,
\nonumber \\
q^{}_\beta \to q^{}_\beta e^{{\rm i} \phi^{}_\beta} \; , &&
q^{}_j \to q^{}_j e^{{\rm i}\phi^{}_j} \; ,
\nonumber \\
q^{}_\gamma \to q^{}_\gamma e^{{\rm i} \phi^{}_\gamma} \; , &&
q^{}_k \to q^{}_k e^{{\rm i}\phi^{}_k} \; , \hspace{0.5cm}
\label{8}
\end{eqnarray}
where $\alpha \neq \beta \neq \gamma$ and $i \neq j \neq k$ run respectively
over $(u, c, t)$ and $(d, s, b)$. In this case we find that the relevant
CKM matrix elements must transform as
\begin{eqnarray}
V^{}_{\alpha i} \to V^{}_{\alpha i}
e^{{\rm i}\left(\phi^{}_\alpha - \phi^{}_i\right)} \; , \quad
V^{}_{\beta j} \to V^{}_{\beta j}
e^{{\rm i}\left(\phi^{}_\beta - \phi^{}_j\right)} \; , \quad
V^{}_{\gamma k} \to V^{}_{\gamma k}
e^{{\rm i}\left(\phi^{}_\gamma - \phi^{}_k\right)} \; ,
\label{9}
\end{eqnarray}
so as to keep the Lagrangian ${\cal L}^{}_{\rm cc}$ of weak charged-current
interactions unchanged
\footnote{All the other parts of the electroweak theory are insensitive to
the above quark phase transformations.}.
As $\det V$ is a linear combination of six ``trios" of the CKM matrix elements
involving six different flavor indices, we immediately arrive at
\begin{eqnarray}
\det V \to e^{{\rm i} \left(\phi^{}_\alpha + \phi^{}_\beta +
\phi^{}_\gamma\right)} \left(\det V\right)
e^{-{\rm i} \left(\phi^{}_i + \phi^{}_j + \phi^{}_k\right)} \;
\label{10}
\end{eqnarray}
corresponding to the phase transformations in Eq.~(\ref{9}). It is therefore
obvious that each of $\lozenge^{ijk}_{\alpha\beta\gamma}$ keeps invariant
under the arbitrary quark phase transformations made in Eq.~(\ref{8}).

We proceed to demonstrate that the imaginary parts of all the six
$\lozenge^{ijk}_{\alpha\beta\gamma}$ in Eq.~(\ref{5}) are equal to
$-{\cal J}$. Taking account of $|\det V| = 1$ for the unitary CKM matrix
$V$, we find
\begin{eqnarray}
\lozenge^{ijk}_{\alpha\beta\gamma} =
V^{}_{\alpha i} V^{}_{\beta j} V^{}_{\gamma k} \left(\det V\right)^*
\hspace{-0.2cm} & = & \hspace{-0.2cm}
\left[|V^{}_{\alpha i}|^2 |V^{}_{\beta j}|^2 - V^{}_{\alpha i} V^{}_{\beta j}
V^*_{\alpha j} V^*_{\beta i}\right] \epsilon^{}_{\alpha\beta\gamma}
\epsilon^{}_{ijk} \; ,
\nonumber \\
\hspace{-0.2cm} & = & \hspace{-0.2cm}
\left[|V^{}_{\alpha i}|^2 |V^{}_{\gamma k}|^2 - V^{}_{\alpha i} V^{}_{\gamma k}
V^*_{\alpha k} V^*_{\gamma i}\right] \epsilon^{}_{\alpha\beta\gamma}
\epsilon^{}_{ijk} \; ,
\nonumber \\
\hspace{-0.2cm} & = & \hspace{-0.2cm}
\left[|V^{}_{\beta j}|^2 |V^{}_{\gamma k}|^2 - V^{}_{\beta j} V^{}_{\gamma k}
V^*_{\beta k} V^*_{\gamma j}\right] \epsilon^{}_{\alpha\beta\gamma}
\epsilon^{}_{ijk} \;  \hspace{0.7cm}
\label{11}
\end{eqnarray}
as a general result. Then a comparison between Eqs.~(\ref{4}) and (\ref{11}) leads
us to ${\rm Im} \lozenge^{ijk}_{\alpha\beta\gamma} = -{\cal J}$.
In other words, both the ``trio" and ``quartet" roads lead to ``Rome" ---
the unique and universal rephasing invariant $\cal J$ of weak CP violation.
This proof tells us that we may also define a set of physical KM phases as
follows~\cite{Xing:2025fpb}:
\begin{eqnarray}
\phi^{ijk}_{\alpha\beta\gamma} \equiv \arctan\left(\frac{\cal J}{
{\rm Re}\lozenge^{ijk}_{\alpha\beta\gamma}}\right) \;
\label{12}
\end{eqnarray}
with $\alpha \neq \beta \neq \gamma$ and $i \neq j \neq k$ for the up-
and down-type quarks. Of course, ${\cal J} \propto
\sin\phi^{ijk}_{\alpha\beta\gamma}$ must hold, just as ${\cal J} \propto
\sin\delta^{}_{\rm KM}$ or $\sin\beta^{}_{\rm KM}$.

To have a ball-park feeling about the sizes of $\phi^{ijk}_{\alpha\beta\gamma}$,
let us calculate six distinctive $\lozenge^{ijk}_{\alpha\beta\gamma}$ with the
help of a sufficiently accurate version~\cite{Charles:2004jd} of the original
Wolfenstein parametrization of $V$~\cite{Wolfenstein:1983yz}. We obtain the
following leading-order results:
\begin{eqnarray}
\phi^{dsb}_{uct} \hspace{-0.2cm} & \simeq & \hspace{-0.2cm}
\arctan\left(A^2 \lambda^6 \eta\right) \simeq 0.0018^\circ \; , \hspace{0.5cm}
\nonumber \\
\phi^{dbs}_{uct} \hspace{-0.2cm} & \simeq & \hspace{-0.2cm}
\arctan\left(-\lambda^2 \eta\right) \simeq 178.98^\circ \; ,
\nonumber \\
\phi^{sdb}_{uct} \hspace{-0.2cm} & \simeq & \hspace{-0.2cm}
\arctan\left(-A^2 \lambda^4 \eta\right) \simeq 179.96^\circ \; ,
\nonumber \\
\phi^{sbd}_{uct} \hspace{-0.2cm} & \simeq & \hspace{-0.2cm}
\arctan\left(\frac{\eta}{1-\rho}\right) \simeq 22.7^\circ \; ,
\nonumber \\
\phi^{bds}_{uct} \hspace{-0.2cm} & \simeq & \hspace{-0.2cm}
\arctan\left(\frac{\eta}{\rho}\right) \simeq 65.7^\circ \; ,
\nonumber \\
\phi^{bsd}_{uct} \hspace{-0.2cm} & \simeq & \hspace{-0.2cm}
\arctan\left(\frac{\eta}{\rho - \rho^2 - \eta^2}\right) \simeq 88.4^\circ \; ,
\hspace{0.5cm}
\label{13}
\end{eqnarray}
where $\lambda \simeq 0.225$, $A \simeq 0.826$, $\rho \simeq 0.159$ and
$\eta \simeq 0.352$~\cite{ParticleDataGroup:2024cfk} have been input. It turns out
that $\phi^{sbd}_{uct} \simeq \beta^{}_{\rm KM}$ holds to a good degree of accuracy.

At this point it is worth mentioning that the so-called {\it relative} rephasing
invariants of the CKM matrix $V$ built from the ``trios" of its nine elements
and the special phase convention $\det V = 1$ have been discussed
in Refs.~\cite{Chang:2002yr,Kuo:2005pf}. Such a phase convention is equivalent
to the requirement $\phi^{}_\alpha + \phi^{}_\beta +
\phi^{}_\gamma = \phi^{}_i + \phi^{}_j + \phi^{}_k$ as can be easily seen from
Eq.~(\ref{10}), and hence the corresponding rephasing invariants only make
{\it relative} sense. Note that the standard parametrization of the CKM matrix $V$
happens to satisfy $\det V = 1$, allowing us to calculate our {\it full}
rephasing invariants $\lozenge^{ijk}_{\alpha\beta\gamma}$ in terms of three
Euler-like angles $\vartheta^{}_{ij}$ (for $ij = 12, 13, 23$) and the KM phase
$\delta^{}_{\rm KM}$ without loss of any generality.

It is also worth mentioning that a physical phase parameter
$\delta \equiv \arg\left(V^{}_{ud} V^{}_{us} V^{}_{cb} V^{}_{tb}
V^{*}_{ub} \det V^*\right)$, built from a ``quintet" of the CKM matrix
elements, has recently been proposed~\cite{Yang:2025law,Yang:2025ftl};
and its analog in the lepton sector has also been discussed by assuming the
PMNS lepton flavor mixing matrix $U$ to be exactly unitary~\cite{Yang:2025hex}.
The key spirit of such exercises is to assure the chosen phase parameter
to be rephasing-invariant and as close to an important observable quantity of
CP violation as possible. There are certainly many different choices in this
connection, just like many different parametrizations of the CKM matrix $V$.

\section{On the PMNS trios}

Now we turn to the $3\times 3$ PMNS lepton flavor mixing matrix $U$ in the
canonical seesaw framework of neutrino mass generation. As $U$ and $R$ in
Eq.~(\ref{6}) are actually two sub-matrices of a $6 \times 6$ unitary matrix
$\mathbb{U}$ utilized to transform the flavor basis of three active neutrinos
and three sterile neutrinos to their mass basis, they are intrinsically correlated
with each other via both the seesaw relation and the unitarity condition.
Considering a complete block parametrization of $\mathbb{U}$ in terms of
the Euler-like rotation angles and phase angles, as first advocated in
Refs.~\cite{Xing:2007zj,Xing:2011ur}, we naturally arrive at $U = A U^{}_0$ with
$A$ being a $3 \times 3$ lower triangular matrix and $U^{}_0$ being a
$3 \times 3$ unitary matrix. The explicit expressions of $A$, $R$
and $U^{}_0$ are given by~\cite{Xing:2011ur}:
\begin{eqnarray}
A \hspace{-0.2cm} & = & \hspace{-0.2cm}
\left( \begin{matrix} c^{}_{14} c^{}_{15} c^{}_{16} & 0 & 0
\cr \vspace{-0.45cm} \cr
\begin{array}{l} -c^{}_{14} c^{}_{15} \hat{s}^{}_{16} \hat{s}^*_{26} -
c^{}_{14} \hat{s}^{}_{15} \hat{s}^*_{25} c^{}_{26} \\
-\hat{s}^{}_{14} \hat{s}^*_{24} c^{}_{25} c^{}_{26} \end{array} &
c^{}_{24} c^{}_{25} c^{}_{26} & 0 \cr \vspace{-0.45cm} \cr
\begin{array}{l} -c^{}_{14} c^{}_{15} \hat{s}^{}_{16} c^{}_{26} \hat{s}^*_{36}
+ c^{}_{14} \hat{s}^{}_{15} \hat{s}^*_{25} \hat{s}^{}_{26} \hat{s}^*_{36} \\
- c^{}_{14} \hat{s}^{}_{15} c^{}_{25} \hat{s}^*_{35} c^{}_{36} +
\hat{s}^{}_{14} \hat{s}^*_{24} c^{}_{25} \hat{s}^{}_{26}
\hat{s}^*_{36} \\
+ \hat{s}^{}_{14} \hat{s}^*_{24} \hat{s}^{}_{25} \hat{s}^*_{35}
c^{}_{36} - \hat{s}^{}_{14} c^{}_{24} \hat{s}^*_{34} c^{}_{35}
c^{}_{36} \end{array} &
\begin{array}{l} -c^{}_{24} c^{}_{25} \hat{s}^{}_{26} \hat{s}^*_{36} -
c^{}_{24} \hat{s}^{}_{25} \hat{s}^*_{35} c^{}_{36} \\
-\hat{s}^{}_{24} \hat{s}^*_{34} c^{}_{35} c^{}_{36} \end{array} &
c^{}_{34} c^{}_{35} c^{}_{36} \cr \end{matrix} \right) \; , \hspace{0.5cm}
\nonumber \\
R \hspace{-0.2cm} & = & \hspace{-0.2cm}
\left( \begin{matrix} \hat{s}^*_{14} c^{}_{15} c^{}_{16} &
\hat{s}^*_{15} c^{}_{16} & \hat{s}^*_{16} \cr \vspace{-0.45cm} \cr
\begin{array}{l} -\hat{s}^*_{14} c^{}_{15} \hat{s}^{}_{16} \hat{s}^*_{26} -
\hat{s}^*_{14} \hat{s}^{}_{15} \hat{s}^*_{25} c^{}_{26} \\
+ c^{}_{14} \hat{s}^*_{24} c^{}_{25} c^{}_{26} \end{array} & -
\hat{s}^*_{15} \hat{s}^{}_{16} \hat{s}^*_{26} + c^{}_{15}
\hat{s}^*_{25} c^{}_{26} & c^{}_{16} \hat{s}^*_{26} \cr \vspace{-0.45cm} \cr
\begin{array}{l} -\hat{s}^*_{14} c^{}_{15} \hat{s}^{}_{16} c^{}_{26}
\hat{s}^*_{36} + \hat{s}^*_{14} \hat{s}^{}_{15} \hat{s}^*_{25}
\hat{s}^{}_{26} \hat{s}^*_{36} \\ - \hat{s}^*_{14} \hat{s}^{}_{15}
c^{}_{25} \hat{s}^*_{35} c^{}_{36} - c^{}_{14} \hat{s}^*_{24}
c^{}_{25} \hat{s}^{}_{26}
\hat{s}^*_{36} \\
- c^{}_{14} \hat{s}^*_{24} \hat{s}^{}_{25} \hat{s}^*_{35}
c^{}_{36} + c^{}_{14} c^{}_{24} \hat{s}^*_{34} c^{}_{35} c^{}_{36}
\end{array} &
\begin{array}{l} -\hat{s}^*_{15} \hat{s}^{}_{16} c^{}_{26} \hat{s}^*_{36}
- c^{}_{15} \hat{s}^*_{25} \hat{s}^{}_{26} \hat{s}^*_{36} \\
+c^{}_{15} c^{}_{25} \hat{s}^*_{35} c^{}_{36} \end{array} &
c^{}_{16} c^{}_{26} \hat{s}^*_{36} \cr \end{matrix} \right) \;
\hspace{0.5cm}
\label{14}
\end{eqnarray}
with $c^{}_{ij} \equiv \cos\theta^{}_{ij}$, $s^{}_{ij} \equiv \sin\theta^{}_{ij}$
and $\hat{s}^{}_{ij} \equiv s^{}_{ij} e^{{\rm i}\delta^{}_{ij}}$ (for $i = 1, 2, 3$
and $j = 4, 5, 6$); and
\begin{eqnarray}
U^{}_0 = \left( \begin{matrix} c^{}_{12} c^{}_{13} & \hat{s}^*_{12}
c^{}_{13} & \hat{s}^*_{13} \cr
-\hat{s}^{}_{12} c^{}_{23} -
c^{}_{12} \hat{s}^{}_{13} \hat{s}^*_{23} & c^{}_{12} c^{}_{23} -
\hat{s}^*_{12} \hat{s}^{}_{13} \hat{s}^*_{23} & c^{}_{13}
\hat{s}^*_{23} \cr
\hat{s}^{}_{12} \hat{s}^{}_{23} - c^{}_{12}
\hat{s}^{}_{13} c^{}_{23} & -c^{}_{12} \hat{s}^{}_{23} -
\hat{s}^*_{12} \hat{s}^{}_{13} c^{}_{23} & c^{}_{13} c^{}_{23}
\cr \end{matrix} \right)
\label{15}
\end{eqnarray}
is a leptonic analog of the CKM matrix $V$ but consists of three CP-violating
phases. A recent global analysis of current electroweak precision measurements,
quark flavor data and neutrino oscillation data has offered some stringent
limits on the deviation of $U U^\dagger = A A^\dagger = {\bf 1} -
R R^\dagger$ from the identity matrix $\bf 1$~\cite{Blennow:2023mqx,Xing:2024gmy}.
Namely, $s^{}_{1j} < 0.051$, $s^{}_{2j} < 0.0047$ and $s^{}_{3j} < 0.045$
in the $m^{}_1 < m^{}_2 < m^{}_3$ case; or
$s^{}_{1j} < 0.053$, $s^{}_{2j} < 0.0045$ and $s^{}_{3j} < 0.040$
in the $m^{}_3 < m^{}_1 < m^{}_2$ (for $j = 4, 5, 6$)~\cite{Xing:2024gmy}.
We are therefore left with
\begin{eqnarray}
A \simeq {\bf 1} - \left( \begin{matrix} a^{}_{11} & 0 & 0 \cr
a^{}_{21} & a^{}_{22} & 0 \cr
a^{}_{31} & a^{}_{32} & a^{}_{33} \cr \end{matrix}
\right) \; ,
\quad\quad
R \simeq \left( \begin{matrix} \hat{s}^*_{14} &
\hat{s}^*_{15} & \hat{s}^*_{16} \cr
\hat{s}^*_{24} & \hat{s}^*_{25} &
\hat{s}^*_{26} \cr
\hat{s}^*_{34} & \hat{s}^*_{35} &
\hat{s}^*_{36} \cr \end{matrix} \right) \; ,
\label{16}
\end{eqnarray}
as two good approximations with
$a^{}_{ii} \equiv \left(s^2_{i4} + s^2_{i 5} + s^2_{i 6}\right)/2$ and
$a^{}_{ij} \equiv \hat{s}^*_{i 4} \hat{s}^{}_{j 4} +
\hat{s}^*_{i 5} \hat{s}^{}_{j 5} + \hat{s}^*_{i 6} \hat{s}^{}_{j 6}$
(for $ii = 11, 22, 33$ and $ij = 21, 31, 32$). It is then straightforward to
write out the PMNS matrix $U = A U^{}_0$ in terms of the parameters appearing
in Eqs.~(\ref{15}) and (\ref{16}).

To calculate the new PMNS rephasing invariants $\blacklozenge^{ijk}_{\alpha\beta\gamma}$
defined in Eq.~({7}), the first step is to figure out the determinant of $U$ as
follows:
\begin{eqnarray}
\det U = \det A \cdot \det U^{}_0 = \prod^3_{i=1} c^{}_{i4} c^{}_{i5} c^{}_{i6}
\simeq 1 - a^{}_{11} - a^{}_{22} - a^{}_{33}
\label{17}
\end{eqnarray}
in the Euler-like parametrization chosen above. Keeping only the non-unitarity
corrections of ${\cal O}(a^{}_{ij})$ (for $ij = 21, 31, 32$), we obtain the
analytical expressions
\begin{eqnarray}
\blacklozenge^{123}_{e\mu\tau}
\hspace{-0.2cm} & \simeq & \hspace{-0.2cm}
c^{}_{12} c^{}_{23} c^{2}_{13} \left( c^{}_{12} c^{}_{23} - s^{}_{12} s^{}_{23}
s^{}_{13} e^{{\rm i}\delta^{}_\nu} \right) - a^{\prime}_{21}
s^{}_{12} c^{}_{12} c^{}_{23} c^{3}_{13}
+ a^{\prime}_{31} c^{}_{12} s^{}_{13} c^{}_{13} \left( s^{}_{12} s^{}_{23} s^{}_{13}
- c^{}_{12} c^{}_{23} e^{-{\rm i}\delta^{}_\nu} \right)
\nonumber \\
\hspace{-0.2cm} & & \hspace{-0.2cm}
- \hspace{0.08cm} a^{\prime}_{32} c^{}_{12} s^{}_{23} c^{2}_{13}
\left( c^{}_{12} c^{}_{23} - s^{}_{12}
s^{}_{23} s^{}_{13} e^{{\rm i}\delta^{}_\nu} \right) \; , \hspace{0.5cm}
\nonumber \\
\blacklozenge^{132}_{e\mu\tau}
\hspace{-0.2cm} & \simeq & \hspace{-0.2cm}
- \hspace{0.08cm} c^{}_{12} s^{}_{23} c^{2}_{13} \left( c^{}_{12} s^{}_{23}
+ s^{}_{12} c^{}_{23} s^{}_{13} e^{{\rm i}\delta^{}_\nu} \right)
+ a^{\prime}_{21} c^{}_{12} s^{}_{13} c^{}_{13} \left( s^{}_{12} c^{}_{23} s^{}_{13}
+ c^{}_{12} s^{}_{23} e^{-{\rm i}\delta^{}_\nu} \right)
\nonumber \\
\hspace{-0.2cm} & & \hspace{-0.2cm}
- \hspace{0.08cm} a^{\prime}_{31} s^{}_{12} c^{}_{12} s^{}_{23} c^{3}_{13}
- a^{\prime}_{32} c^{}_{12} s^{}_{23} c^{2}_{13} \left( c^{}_{12} c^{}_{23} -
s^{}_{12} s^{}_{23} s^{}_{13} e^{{\rm i}\delta^{}_\nu} \right) \; , \hspace{0.5cm}
\nonumber \\
\blacklozenge^{213}_{e\mu\tau}
\hspace{-0.2cm} & \simeq & \hspace{-0.2cm}
- \hspace{0.08cm} s^{}_{12} c^{}_{23} c^{2}_{13} \left( s^{}_{12} c^{}_{23}
+ c^{}_{12} s^{}_{23} s^{}_{13} e^{{\rm i}\delta^{}_\nu} \right)
- a^{\prime}_{21} s^{}_{12} c^{}_{12} c^{}_{23} c^{3}_{13}
+ a^{\prime}_{31} s^{}_{12} s^{}_{13} c^{}_{13} \left( c^{}_{12} s^{}_{23} s^{}_{13}
+ s^{}_{12} c^{}_{23} e^{-{\rm i}\delta^{}_\nu} \right)
\nonumber \\
\hspace{-0.2cm} & & \hspace{-0.2cm}
+ \hspace{0.08cm} a^{\prime}_{32}s^{}_{12} s^{}_{23} c^{2}_{13} \left( s^{}_{12} c^{}_{23}
+ c^{}_{12} s^{}_{23} s^{}_{13} e^{{\rm i}\delta^{}_\nu} \right) \; , \hspace{0.5cm}
\nonumber \\
\blacklozenge^{231}_{e\mu\tau}
\hspace{-0.2cm} & \simeq & \hspace{-0.2cm}
s^{}_{12} s^{}_{23} c^{2}_{13} \left( s^{}_{12} s^{}_{23} - c^{}_{12} c^{}_{23}
s^{}_{13} e^{{\rm i}\delta^{}_\nu} \right)
+ a^{\prime}_{21} s^{}_{12} s^{}_{13} c^{}_{13} \left( c^{}_{12} c^{}_{23} s^{}_{13}
- s^{}_{12} s^{}_{23} e^{-{\rm i}\delta^{}_\nu} \right)
\nonumber \\
\hspace{-0.2cm} & & \hspace{-0.2cm}
- \hspace{0.08cm} a^{\prime}_{31} s^{}_{12} c^{}_{12} s^{}_{23} c^{3}_{13}
+ a^{\prime}_{32} s^{}_{12} s^{}_{23} c^{2}_{13} \left( s^{}_{12} c^{}_{23}
+ c^{}_{12} s^{}_{23} s^{}_{13} e^{{\rm i}\delta^{}_\nu} \right) \; , \hspace{0.5cm}
\nonumber \\
\blacklozenge^{312}_{e\mu\tau}
\hspace{-0.2cm} & \simeq & \hspace{-0.2cm}
\left( c^{2}_{12} s^{2}_{23} + s^{2}_{12} c^{2}_{23} \right) s^{2}_{13} +
s^{}_{12} c^{}_{12} s^{}_{23} c^{}_{23} s^{}_{13} \left[ \left(1 + s^{2}_{13} \right)
\cos\delta^{}_\nu - {\rm i} \hspace{0.05cm} c^{2}_{13} \sin\delta^{}_\nu \right]
\nonumber \\
\hspace{-0.2cm} & & \hspace{-0.2cm}
+ \hspace{0.08cm} a^{\prime}_{21} c^{}_{12} s^{}_{13} c^{}_{13} \left( s^{}_{12}
c^{}_{23} s^{}_{13} + c^{}_{12} s^{}_{23} e^{-{\rm i}\delta^{}_\nu} \right)
+ a^{\prime}_{31} s^{}_{12} s^{}_{13} c^{}_{13} \left( c^{}_{12}
s^{}_{23} s^{}_{13} + s^{}_{12} c^{}_{23} e^{-{\rm i}\delta^{}_\nu} \right)
\nonumber \\
\hspace{-0.2cm} & & \hspace{-0.2cm}
+ \hspace{0.08cm} a^{\prime}_{32} \left[ s^{}_{23} c^{}_{23} s^{2}_{13} \cos2\theta^{}_{12} +
s^{}_{12} c^{}_{12} s^{}_{13} \left( c^{2}_{23} e^{-{\rm i}\delta^{}_\nu} -
s^{2}_{23} s^{2}_{13} e^{{\rm i}\delta^{}_\nu} \right) \right] \; , \hspace{0.5cm}
\nonumber \\
\blacklozenge^{321}_{e\mu\tau}
\hspace{-0.2cm} & \simeq & \hspace{-0.2cm}
- \left( s^{2}_{12} s^{2}_{23} + c^{2}_{12} c^{2}_{23} \right)
s^{2}_{13} + s^{}_{12} c^{}_{12} s^{}_{23} c^{}_{23} s^{}_{13} \left[ \left(1 +
s^{2}_{13} \right) \cos\delta^{}_\nu - {\rm i} \hspace{0.05cm} c^{2}_{13} \sin\delta^{}_\nu \right]
\nonumber \\
\hspace{-0.2cm} & & \hspace{-0.2cm}
+ \hspace{0.08cm} a^{\prime}_{21} s^{}_{12} s^{}_{13} c^{}_{13} \left( c^{}_{12} c^{}_{23}
s^{}_{13} - s^{}_{12} s^{}_{23} e^{-{\rm i}\delta^{}_\nu} \right)
+ a^{\prime}_{31} c^{}_{12} s^{}_{13} c^{}_{13} \left( s^{}_{12} s^{}_{23} s^{}_{13} -
c^{}_{12} c^{}_{23} e^{-{\rm i}\delta^{}_\nu} \right)
\nonumber \\
\hspace{-0.2cm} & & \hspace{-0.2cm}
+ \hspace{0.08cm} a^{\prime}_{32} \left[ s^{}_{23} c^{}_{23} s^{2}_{13} \cos2\theta^{}_{12} +
s^{}_{12} c^{}_{12} s^{}_{13} \left( c^{2}_{23} e^{-{\rm i}\delta^{}_\nu} -
s^{2}_{23} s^{2}_{13} e^{{\rm i}\delta^{}_\nu} \right) \right] \; , \hspace{0.5cm}
\label{18}
\end{eqnarray}
where $\delta^{}_\nu \equiv \delta^{}_{13} - \delta^{}_{12} - \delta^{}_{23}$ is 
a KM-like phase~\cite{Kobayashi:1977ss}, and $a^\prime_{21} \equiv a^{}_{21} 
e^{-{\rm i}\delta^{}_{12}}$, $a^\prime_{31} \equiv a^{}_{31} e^{-{\rm i}
\left(\delta^{}_{12} + \delta^{}_{23}\right)}$
and $a^\prime_{32} \equiv a^{}_{32} e^{-{\rm i}\delta^{}_{23}}$ are defined
\footnote{Note that the parameters of $D^{}_\nu$ and $U^{}_0$ associated with
three light Majorana neutrinos can actually be derived from the parameters of
$D^{}_N$ and $R$ associated with three heavy Majorana neutrinos and
active-sterile flavor mixing effects via the seesaw relation $U D^{}_\nu U^T =
- R D^{}_N R^T$, but the relevant analytical results~\cite{Xing:2024gmy} and
the one-loop renormalization-group running effects~\cite{Huang:2025ubs} are
both very complicated. For the sake of simplicity, we choose not
to take into account the intrinsic (seesaw-induced) correlations between these
two sets of flavor parameters in the present work.}.
Then we arrive at 
${\rm Im} \blacklozenge^{ijk}_{\alpha\beta\gamma} \simeq -{\cal J}^{}_\nu$
with ${\cal J}^{}_\nu = c^{}_{12} s^{}_{12} c^2_{13} s^{}_{13} c^{}_{23} s^{}_{23}
\sin\delta^{}_\nu$ in the leading-order approximation. The 
next-to-leading-order terms of ${\rm Im} \blacklozenge^{ijk}_{\alpha\beta\gamma}$
are suppressed by the smallness of $|a^{}_{ij}|$ of at most ${\cal O}(10^{-3})$, 
but their expressions are not the same as those obtained from the ``quartets" 
of the PMNS matrix $U$~\cite{Xing:2020ivm,Xing:2025zqt}. Given that the T2K 
experiment hints at $|{\cal J}^{}_\nu| \sim {\cal O}(10^{-2})$ to a large
extent~\cite{T2K:2023smv,T2K:2024wfn}, the non-unitarity effects on leptonic
CP violation in neutrino oscillations are certainly negligible~\cite{Xing:2024xwb}.  

\section{Summary and remarks}

In this work we have proposed a new set of rephasing invariants 
$\lozenge^{ijk}_{\alpha\beta\gamma}$ in terms of the ``trios" of nine 
elements of the CKM quark flavor mixing matrix $V$, and demonstrated ${\rm Im}
\lozenge^{ijk}_{\alpha\beta\gamma} = - {\cal J}$ with $\cal J$ being 
the famous Jarlskog invariant of CP violation which is built from the
CKM ``quartets". 
Similar rephasing invariants $\blacklozenge^{ijk}_{\alpha\beta\gamma}$ have 
been proposed for the $3\times 3$ PMNS lepton flavor mixing matrix $U$, and
${\rm Im} \blacklozenge^{ijk}_{\alpha\beta\gamma} \simeq -{\cal J}^{}_\nu$ 
has been proved in the canonical seesaw mechanism with ${\cal J}^{}_\nu$ 
being the leptonic Jarlskog invariant in the unitarity limit of $U$.

We find that it is conceptually interesting to study CP violation by
considering those CKM or PMNS ``trios", because they involve
all the six flavor indices of quark or lepton fields and thus reflect
the three-family feature of CP violation based on the standard KM mechanism.
In comparison, one may wonder why ${\cal J} = {\rm Im}\left(V^{}_{ud}
V^{}_{cs} V^*_{us} V^*_{cd}\right)$ is also a correct measure of CP
violation although it only involves four different quark flavors
which belong to the $2\times 2$ Cabibbo matrix. The reason is simply 
that the Cabibbo matrix is actually a submatrix of the $3\times 3$ 
CKM matrix $V$, and hence it is not exactly unitary in the three-family
flavor mixing scheme and can then accommodate CP violation. That is 
why the PMNS matrix $U$ may contain some extra CP-violating effects 
originating from its slight non-unitarity in the canonical seesaw framework. 

The CKM or PMNS ``trio" description of rephasing invariants discussed 
in this paper can certainly be generalized to the case for an $n \times n$ 
unitary flavor mixing matrix and CP violation. But whether some useful
applications can be found in this connection remains an open question.
Further studies along this line of thought should be worthwhile.

\section*{Acknowledgements}

This research was supported in part by the Scientific and Technological 
Innovation Program of the Institute of High Energy Physics under Grant 
No. E55457U2.

\bibliographystyle{elsarticle-num}
\bibliography{Invariant}

\end{document}